\documentclass[aps,twocolumn,pra,superscriptaddress,amsmath,amssymb]{revtex4-1}
\ifx\pdfoutput\undefined
\usepackage[dvipdfmx]{graphicx}
\usepackage[dvipdfmx]{hyperref}
\else
\usepackage{graphicx}
\usepackage{hyperref}
\fi

\usepackage{txfonts}
\usepackage[T1]{fontenc}
\usepackage{xspace}

\usepackage{dcolumn}
\usepackage{bm}
\usepackage{amsmath}
\usepackage{booktabs}
\usepackage[usenames]{color}
\usepackage{xcolor}

\usepackage{mathrsfs}

\usepackage{ulem}

\usepackage{soul}
\soulregister\cite7
\soulregister\ref7
\soulregister\pageref7

\DeclareRobustCommand{\nchange}[2]{\ifmmode{{\textrm{\setul{}{1pt}\setstcolor{blue}\st{$\displaystyle#1$}}}}\else{{\setul{}{1pt}\setstcolor{blue}\st{#1}}}\fi\ \textcolor{blue}{#2}}
\begin{document}
\let\emph\textit

\title{
Ferromagnetic Instability for single-band Hubbard model in the strong-coupling regime
}

\author{Yusuke Kamogawa}
\affiliation{Department of Physics, Tokyo Institute of Technology, Meguro, Tokyo 152-8551, Japan}
\author{Joji Nasu}
\affiliation{Department of Physics, Tokyo Institute of Technology, Meguro, Tokyo 152-8551, Japan}
\affiliation{
  Department of Physics, Yokohama National University,
  79-5 Tokiwadai, Hodogaya, Yokohama 240-8501, Japan
}

\author{Akihisa Koga}
\affiliation{Department of Physics, Tokyo Institute of Technology, Meguro, Tokyo 152-8551, Japan}

 \date{\today}
\begin{abstract}
  We study a ferromagnetic instability in a doped single-band Hubbard model
  by means of dynamical mean-field theory with
  the continuous-time quantum Monte Carlo simulations.
  Examining the effect of the strong correlations in the system
  on the hypercubic and Bethe lattice,
  we find that the ferromagnetically ordered state appears in the former,
  while it does not in the latter.
  We also reveal that the ferromagnetic order is more stable in the case that
  the noninteracting DOS exhibits a slower decay in the high-energy region.
  The present results suggest that, in the strong-coupling regime,
  the high-energy part of DOS plays an essential role
  for the emergence of the ferromagnetically ordered state, in contrast to the Stoner criterion
  justified in the weak interaction limit.
\end{abstract}

\maketitle
%%%%%%%%%%%%%%%%% 1st section %%%%%%%%%%%%%%%%%%%%%
\section{Introduction}
%%%%%%%%%%%%%%%%%%%%%%%%%%%%%%%%%%%%%%%%%%%%%%%%%%%

Ferromagnetic (FM) metallic state in the strongly correlated electron systems
is a long standing problem though iron is known to be a magnet from ancient times.
In the multiorbital system,
there exists the Hund coupling between electrons in degenerate orbitals,
which tends to realize the FM ordered states
at low temperatures~\cite{Slater1,Slater2,Zener,VanVleck}.
In fact, the ordered state has been reported in the doped Hubbard model with
degenerate orbitals~\cite{Momoi},
and double exchange model~\cite{Motome},
which should be relevant for the realistic materials
such as $\rm La_{1-x}Sr_xMnO_3$~\cite{Tokura}.
By contrast, the FM instability in simpler models
is less understood.
When the Hartree approximation is applied to the single-band Hubbard model,
one meets the Stoner criterion, namely, the FM instability
appears due to the Coulomb interaction when the system has
a large density of states (DOS) at the Fermi level.
This criterion is qualitatively correct in the weak coupling region.
In fact, the existence of the FM ordered states has been clarified
in the single band systems
with flat bands~\cite{Mielke1,Mielke2,Mielke3,Tasaki,Kusakabe,Miyahara,Noda}
and asymmetric DOS~\cite{Kanamori,Ulmke,Wahle,BalzerPotthoff}.

In the case with strong Coulomb interactions, the Stoner theory is not applicable because of the large modulation of the low-temperature susceptibility.
To take into account the spin fluctuations, theoretical attempts
%such as self-consistent renormalization theory
have been devoted so far~\cite{Murata,Moriya1,Moriya2}.
In the strong-coupling limit, intersite correlations
via the effective Heisenberg interactions
should be dominant, which enhances antiferromagnetic (AFM) fluctuations
against the FM instability.
In the bipartite system in the $d>1$ dimensions,
the AFM ordered state is always realized at half filling~\cite{Penrose}.
Away from half filling, doped holes should gain the kinetic energy, and
therefore it is not trivial that the AFM ordered state survives
in the strong-coupling limit.
On the other hand,
Nagaoka has proved that for a single hole
in the Hubbard model on a lattice with closed loops,
the ground state is a fully polarized ferromagnet in this limit,
so-called Nagaoka ferromagnetism~\cite{Nagaoka}.
Therefore, it is still controversial how stable
such a polarized ordered state
is in the system with the finite hole density.
An important point is that, in this strong-coupling region,
large Coulomb interactions and low energy metallic properties are necessary to
take into account precisely in an equal footing.
In our paper,
we use dynamical mean-field theory (DMFT)~\cite{Metzner,Muller,RevModPhys.68.13,Pruschke},
which is one of the appropriate frameworks to take into account the wide range of energy scales,
to discuss the FM instability in the infinite dimensions.
It has already been clarified that the asymmetry of the DOS plays
an important role for realizing the FM state
in the weak coupling region~\cite{Ulmke,Wahle,BalzerPotthoff}.
As for the strong-coupling region, it has been clarified that
the FM ordered state is not realized in the system
on the Bethe lattice~\cite{Peters,BalzerPotthoff},
but in the hypercubic lattice~\cite{NCA,Zitzler2002}.
These facts in the strong-coupling region
may be understood in terms of the Nagaoka mechanism
since there are no closed loop in the Bethe lattice.
However, in the framework of DMFT, the lattice structure is involved
only via the noninteracting DOS,
which should lead to a minor change in the system, {\it e.g.},
the critical interactions for Mott transitions~\cite{Bulla1999,Potthoff2003}.
Therefore,
key factors for stabilizing the strong-coupling FM ordered state remain unclear.
Furthermore, quantitative treatments are still lacking even in the infinite dimensional systems
since the conventional impurity solvers
such as the noncrossing approximation~\cite{NCA1,NCA2,NCA3,NCA} and
numerical renormalization group~\cite{Krishna,BullaRMP,Sakai,Bulla,Zitzler2002,Peters}
are hard to obtain the dynamical quantities
in both low and extremely high energy regions precisely.
To overcome this,
in this study, we make use of the continuous-time quantum Monte Carlo (CTQMC)
method~\cite{PhysRevLett.97.076405,RevModPhys.83.349}
based on the segment algorithm.
We then discuss the FM instability in the system more precisely
to determine the finite temperature phase diagram.

The paper is organized as follows.
In Sec.~\ref{sec:model},
we introduce the single-band Hubbard model and briefly explain
the framework of DMFT.
In Sec.~\ref{sec:result}, we consider the infinite dimensional
Hubbard model on the hypercubic and Bethe lattices to discuss the FM instability
at low temperatures.
The effect of the noninteracting DOS is also addressed,
by examining magnetic properties in the system with $t$-distribution DOS.
A summary is given in the final section.

%%%%%%%%%%%%%%%%%%%%%%%%%%%%%%%%%%%%%%
\section{Model and methods}\label{sec:model}
%%%%%%%%%%%%%%%%%%%%%%%%%%%%%%%%%%%%%%
We consider the single-band Hubbard model, which
is described by the following Hamiltonian as,
%%%%%%%%%%%%%%%%%%%%%%%%%%%%%%%%%
\begin{align}
	H = & -t\sum_{\langle i,j,\rangle ,\sigma} (c_{i\sigma}^{\dagger} c_{j\sigma} + \mathrm{h.c.})
	+ U \sum_i n_{i\uparrow} n_{i\downarrow}  \nonumber \\
	& - \sum_{i\sigma} (\mu +\frac{h}{2}\sigma) n_{i\sigma},
\end{align}
%%%%%%%%%%%%%%%%%%%%%%%%%%%%%%%%%
where $c_{i\sigma} (c_{i\sigma}^\dag)$ annihilates (creates) an electron
with spin $\sigma(=\uparrow, \downarrow)$ at the $i$th site
and $n_{i\sigma}=c_{i\sigma}^\dag c_{i\sigma}$.
$t$ is the transfer integral, $U$ the on-site interaction,
$\mu$ the chemical potential, and $h$ the external magnetic field.

To study magnetic properties in the single-band Hubbard model,
we make use of DMFT~\cite{Metzner,Muller,RevModPhys.68.13,Pruschke}.
In DMFT, the lattice model is mapped to the problem of a single impurity connected dynamically
to a ``heat bath''.
The electron Green's function is obtained
via the self-consistent solution of this impurity problem.
The treatment is exact in the limit of the infinite dimensions
since nonlocal electron correlations are irrelevant.
The selfenergy is reduced to be site-diagonal $\Sigma_\sigma(k,z)=\Sigma_\sigma(z)$ and
the lattice Green's function is given as
%%%%%%%%%%%%%%%%%%%%%%%%%%%%%%%%%
\begin{eqnarray}
  G_\sigma (k,z)^{-1} &=&   G_{0\sigma} (k,z)^{-1} -\Sigma_\sigma(z),
\end{eqnarray}
%%%%%%%%%%%%%%%%%%%%%%%%%%%%%%%%%
where $G_{0\sigma} (k,z)^{-1}=z+ \mu +\frac{h}{2}\sigma-\epsilon_k$ and
$\epsilon_k$ is the dispersion relation.
The local Green's function is then obtained as
%%%%%%%%%%%%%%%%%%%%%%%%%%%%%%%%%
\begin{eqnarray}
  G_{loc,\sigma} (z) &=&
  \int dk  G_\sigma (k,z)\\
  &=&\int dx \frac{\rho_0(x)}{z+ \mu +\frac{h}{2}\sigma- x - \Sigma_{\sigma} (i \omega_n )},
  \label{rho}
\end{eqnarray}
%%%%%%%%%%%%%%%%%%%%%%%%%%%%%%%%%
where we introduce the non-interacting DOS
$\rho_0(x)=\int dk \delta (x-\epsilon_k)$.
In the effective impurity model,
the Dyson equation is given as,
\begin{eqnarray}
  {\cal G}_\sigma (z)^{-1} &=&   G_{imp, \sigma} (z)^{-1} +\Sigma_{imp, \sigma}(z),
\end{eqnarray}
where ${\cal G}(z)$ is the effective bath.
Solving the effective impurity model,
one can obtain the selfenergy and Green function.
We iterate the selfconsistency conditions
$G_\sigma(z)=G_{imp, \sigma}(z)$ and $\Sigma_\sigma(z)=\Sigma_{imp, \sigma}(z)$
until the desired numerical accuracy is achieved.

In our calculations,
we make use of the hybridization expansion
CTQMC simulations~\cite{PhysRevLett.97.076405,RevModPhys.83.349}
based on the segment algorithm, which is one of the powerful methods
to solve the effective impurity model.
In the method, Monte Carlo samplings are efficiently performed
by local updates such as insertion (removal) of a segment or empty space
between segments (antisegment), or shifts of segment end points.
However, the acceptance probabilities are exponentially suppressed
with respect to the interaction strength $U$.
Therefore, it is hard to evaluate the Green's function
in the reasonable computational cost when $U\gg t$.
Here, we also use additional updates, where
the configuration for both spins in a certain interval
is simultaneously changed.
This allows us to perform the CTQMC method in the strong-coupling region
efficiently~\cite{KogaDBLE}.
Furthermore, we use the intermediate representation
for the Green's function~\cite{PhysRevB.96.035147} since
$G_{imp}(\tau)$ is expected to change rapidly around $\tau=0, \beta$
in the strong-coupling region.

To discuss magnetic properties in the single band Hubbard model,
we calculate the uniform magnetization and magnetic susceptibility,
which are defined as,
\begin{eqnarray}
  m&=&\frac{1}{2}\sum_{i}\Big( \langle n_{i\uparrow}\rangle -\langle n_{i\downarrow} \rangle\Big),\\
  \chi&=&\lim_{h\rightarrow 0}\frac{m}{h}.
\end{eqnarray}
In our calculations, the magnetic susceptibility is numerically evaluated by
the induced magnetization
in the system with a fixed chemical potential
since the modulation of the electron number is confirmed
to be negligible in the presence of $h$.
Here,
we focus on the nature of the FM metallic state
in the single-band Hubbard model.
For this purpose, we neglect the AFM ordered state and phase separation,
which should be realized close to half filling $n\sim 1$~\cite{NCA,Zitzler2002},
where $n=\sum_{i\sigma}\langle n_{i\sigma}\rangle/N$.
This simplification allows us to exhibit the essence of the FM instability
in the strong-coupling region.

\begin{figure}[htb]
  \begin{center}
    \includegraphics[width=\columnwidth]{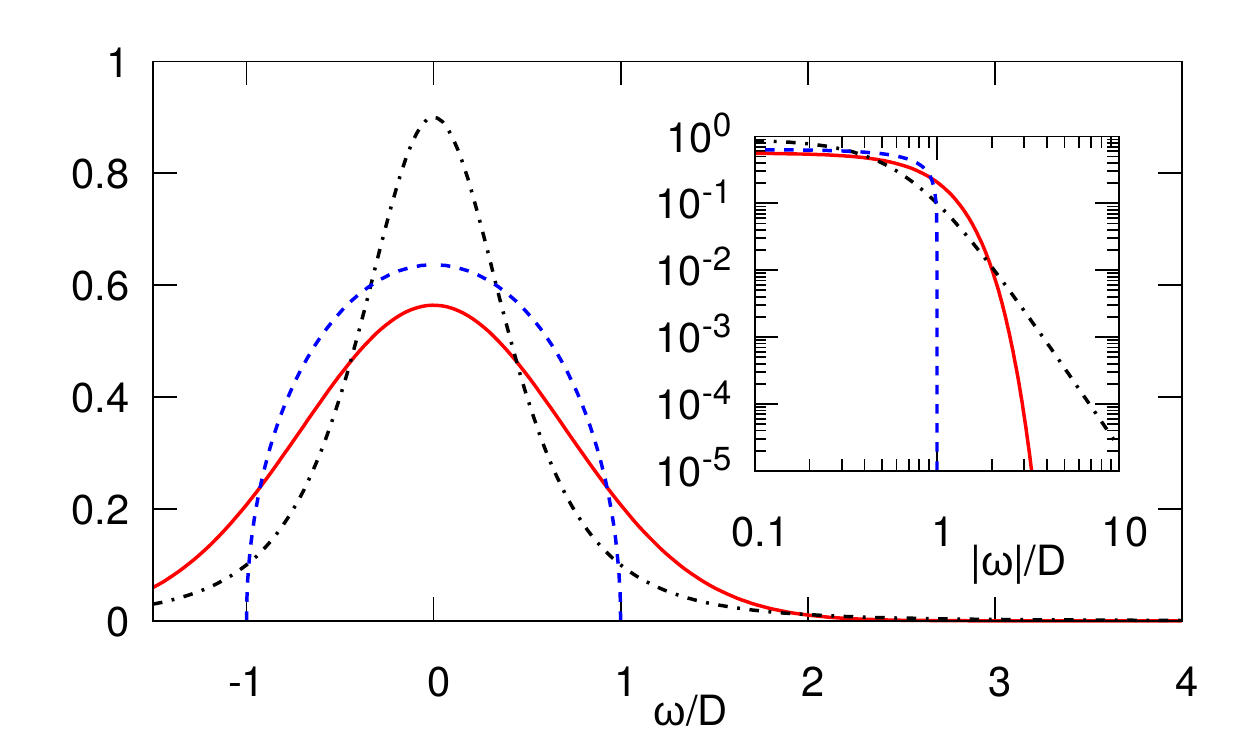}
    \caption{Solid and dashed lines represent DOS in the noninteracting system
      on the hypercubic and Bethe lattices.
      Dot-dashed line represent the $t$-distribution,
      which will be defined in Eq. (\ref{eq:t}).
      The inset shows the tails of their DOS in the large $|\omega|$ region.
    }
    \label{fig:dos}
  \end{center}
\end{figure}

In our paper, we consider the Hubbard model on the Bethe and
hypercubic lattices to study their magnetic properties.
The corresponding noninteracting DOS, which is important in the framework of DMFT
[see Eq. (\ref{rho})],
are given as,
%%%%%%%%%%%%%%%%%%%%%%%%%%%%%%%%%
\begin{eqnarray}
	\rho_{b} (x) &=& \frac{2}{\pi D}\sqrt{1-\left( \frac{x}{D} \right)^2},\\
	\rho_{hc} (x) &=& \frac{1}{\sqrt{\pi} D} \exp\left[-\left(\frac{x}{D}\right)^2\right],\label{eq:1}
\end{eqnarray}
%%%%%%%%%%%%%%%%%%%%%%%%%%%%%%%%%
where $D$ is the characteristic energy scale.
It has been clarified that the difference in the shape of DOS simply
leads to the quantitative change
in the critical interaction of the Mott transition~\cite{Bulla1999,Potthoff2003}.
On the other hand, away from commensurate fillings,
the shape has been discussed to be crucial for the instability
to the FM ordered state in strongly correlated metals~\cite{NCA,Zitzler2002,Peters,BalzerPotthoff}.
In particular, as for the above two forms of the DOS, their high-energy parts are obviously different; the DOS of the hypercubic lattice has a exponentially decaying tail, but that of the Bethe lattice is finite only in the limited region,
as shown in Fig.~\ref{fig:dos}.
%%%%%%%%%%%%%%%%%%%%%%%%%%%%%%%%%
%%%%%%%%%%%%%%%%%%%%%%%%%%%%%%%%%
In the following, we discuss the role of the shape of the DOS for the FM instability
through the systematic finite temperature calculations.

%%%%%%%%%%%%%%%%%%%%%%%%%%%%%%%%%%%%%%
\section{Numerical results}\label{sec:result}
\begin{figure}[htb]
  \begin{center}
    \includegraphics[width=\columnwidth]{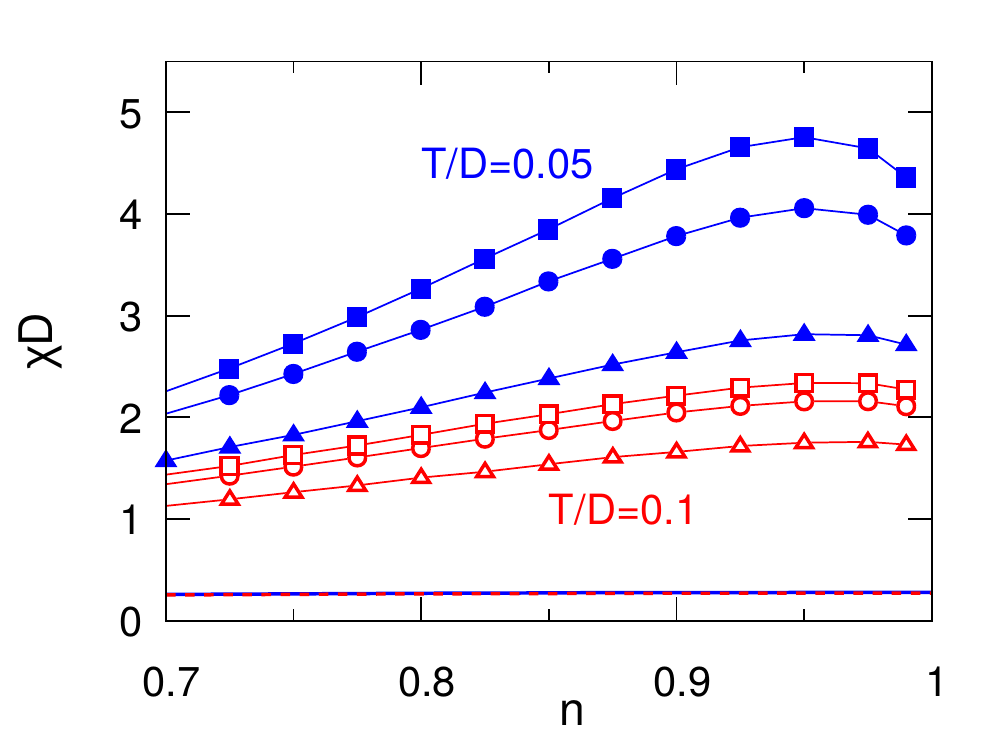}
    \caption{Magnetic Susceptibility as a function of band filling $n$
      in the system on the hypercubic lattice
      when $U/D=10$ (triangles), $20$ (circles) and $30$ (squares)
      at the temperatures $T/D=0.1$ (open symbols) and $0.05$ (solid symbols).
      Solid and dashed lines around $\chi D\sim 0.3$ are the results
      for the noninteracting system at $T/D=0.05$ and $0.1$.
    }
    \label{fig:HC-nchi}
  \end{center}
\end{figure}

%%%%%%%%%%%%%%%%%%%%%%%%%%%%%%%%%%%%%%
We first consider the Hubbard model on the hypercubic lattice~\cite{NCA,Zitzler2002}
to clarify the presence of the FM ordered phase
when the system is away from half filling $(n<1)$ in the case of large Coulomb interactions.
Figure~\ref{fig:HC-nchi} shows the magnetic susceptibility
in the single-band Hubbard model at $T/D=0.05$ and $0.1$.
%%%%%%%%%%%%%%%%%%%%%%%%%%%%%%%%%
%%%%%%%%%%%%%%%%%%%%%%%%%%%%%%%%%
It is found that in the noninteracting system $(U=0)$,
the susceptibility little depends on the electron density and temperature
in this scale $(\chi D\sim 0.3)$.
When the interaction strength is much larger than
the hopping (bandwidth) and temperature,
nonmonotonic behavior appears in the susceptibility as a function of the filling $n$.
The peak structure develops with increasing the Coulomb interaction and decreasing the temperature.
At the low temperature $T/D=0.05$, the susceptibility has a maximum around $n\sim 0.95$, where
ferromagnetic fluctuations are enhanced.
This suggests that
the FM instability appears away from the half filling
when the system has a larger interaction strength
at lower temperatures.

\begin{figure}[htb]
  \begin{center}
    \includegraphics[width=\columnwidth]{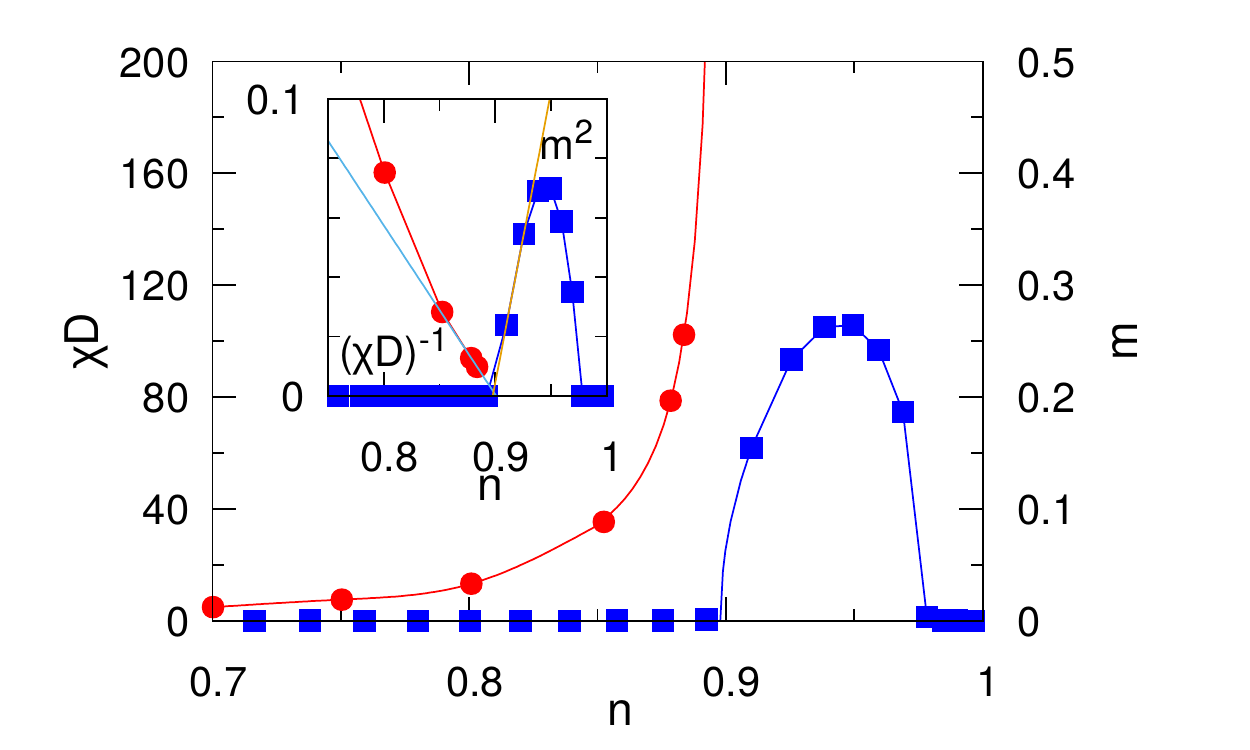}
    \caption{Uniform magnetic susceptibility and magnetization as a function of the electron density
      in the system with $U/D=100$ at the temperature $T/D=0.01$.
      The inset shows critical behavior of the susceptibility and magnetization.
      Solid lines are guides to the eyes.
    }
    \label{fig:HC-nm}
  \end{center}
\end{figure}

To examine the presence of the FM ordered phase at finite interactions and temperatures,
we calculate the uniform susceptibility and magnetization
in the system with $U/D=100$ at $T/D=0.01$, as shown in Fig.~\ref{fig:HC-nm}.
%%%%%%%%%%%%%%%%%%%%%%%%%%%%%%%%%
%%%%%%%%%%%%%%%%%%%%%%%%%%%%%%%%%
In the small $n$ case, the system is in the paramagnetic (PM) state
with the finite susceptibility.
Increasing the electron number, the susceptibility monotonically increases and
at last diverges at the critical value $n=n_{c1}$.
Beyond the critical value, the finite magnetization is induced,
implying that the FM ordered state is realized
in the single-band Hubbard model on the hypercubic lattice.
The magnetization has a maximum around $n\sim 0.95$ and finally
it vanishes at $n=n_{c2}$, where the phase transition
occurs again to the PM metallic state.
By examining critical behavior,
we obtain the critical densities $n_{c1}=0.90$ and $n_{c2}=0.98$.

\begin{figure}[htb]
  \begin{center}
    \includegraphics[width=\columnwidth]{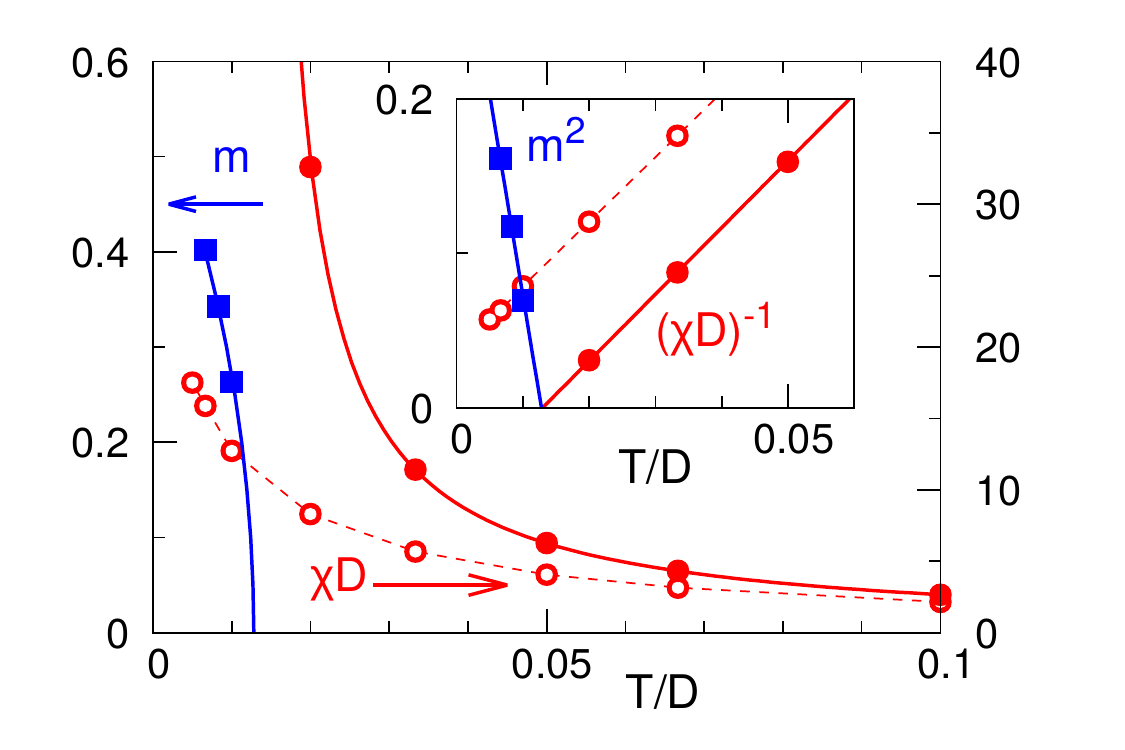}
    \caption{Solid squares (circles) represent magnetization (magnetic susceptibility)
      as a function of the temperature
      in the system on the hypercubic lattice when $U/D=100$ and $n=0.95$.
      Open circles represent the magnetic susceptibility for the system
      with $U/D=20.0$ and $n=0.95$.
      The inset shows critical behavior of these quantities.
      Solid and dashes lines are guides to eyes.
    }
    \label{fig:HC-Tchiinv}
  \end{center}
\end{figure}

To reveal how stable the FM ordered state is against thermal fluctuations,
we show in Fig.~\ref{fig:HC-Tchiinv}
the temperature dependence of the magnetization and magnetic susceptibility
in the system with $U/D=100$ and $n=0.95$.
%%%%%%%%%%%%%%%%%%%%%%%%%%%%%%%%%
%%%%%%%%%%%%%%%%%%%%%%%%%%%%%%%%%
We find that decreasing temperatures, the magnetic susceptibility monotonically increases
and at last, diverges at a finite temperature $T_c$.
Further decrease of temperatures drives the system to
the FM ordered state with the uniform magnetization $m$.
The critical temperature ${T_c}/D\sim 0.013$ is obtained,
examining critical behavior in these quantities
$m\sim (T_c-T)^\beta$ and $\chi \sim (T-T_c)^{-\gamma}$ with $\beta=1/2$ and $\gamma=1$,
as shown in the inset of Fig.~\ref{fig:HC-Tchiinv}.
These critical exponents are consistent with the mean-field theory.
On the other hand, in the case with $U/D=20$,
the magnetic susceptibility approaches a certain value with decreasing temperatures,
implying that the ground state is the PM metal.

\begin{figure}[htb]
  \begin{center}
    \includegraphics[width=\columnwidth]{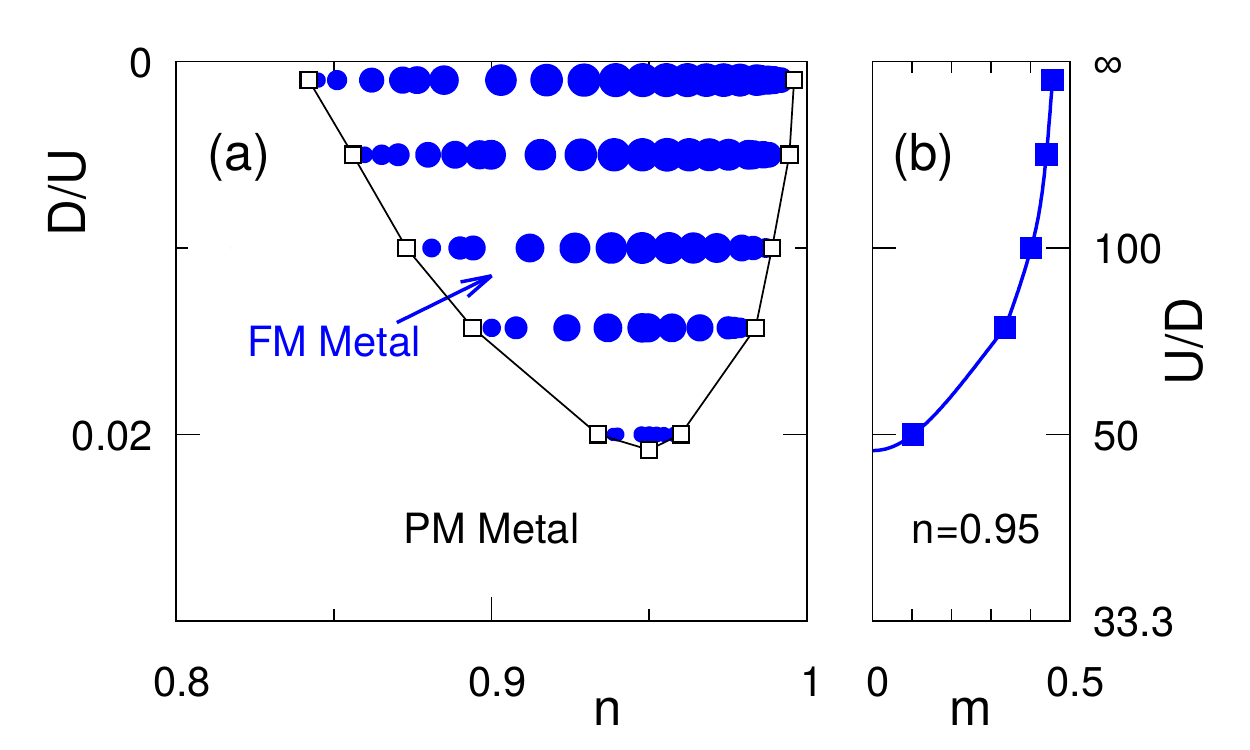}
    \caption{(a) Phase diagram of the single-band Hubbard model on the hypercubic lattice
      at the temperature $T/D=0.0067$.
      Circle area is proportional to the moment size.
      (b) Uniform magnetization as a function of the Coulomb interaction
      in the system with $n=0.95$ at $T/D=0.0067$.
    }
    \label{fig:HC-PD}
  \end{center}
\end{figure}

By performing similar calculations for different values of $U$ and $n$,
we obtain the phase diagram at the temperature $T/D=0.0067$,
as shown in Fig.~\ref{fig:HC-PD}(a).
%%%%%%%%%%%%%%%%%%%%%%%%%%%%%%%%%
%%%%%%%%%%%%%%%%%%%%%%%%%%%%%%%%%
It is found that the FM ordered state is realized around $n\sim 0.95$
in the strong-coupling regime.
In addition, increasing the interaction strength,
the magnetization smoothly increases and approaches a certain value at the fixed temperature,
as shown in Fig.~\ref{fig:HC-PD}(b).
This means that the FM ordered state becomes stable even in the large $U$ region.
This is in contrast to the AFM ordered state at half filling.
In the state, the AFM order parameter decreases with increasing the interaction
at a fixed temperature
since intersite correlations scaled by $\sim t^2/U$ in the strong-coupling
limit~\cite{KogaDBLE,Yanatori}.
By contrast, in the case away from the half filing,
the uniform magnetization is saturated in the large $U$ limit,
as shown in Fig.~\ref{fig:HC-PD}(b).
This suggests that the stability of the FM ordered state is dominated by the kinetic energy.
This is similar to the origin of the Nagaoka ferromagnetism,
implying that the FM ordered state we find is adiabatically connected
to the Nagaoka ferromagnetism, which is justified in the limits of $U\to\infty$ and $n\to 1$.

\begin{figure}[htb]
  \begin{center}
    \includegraphics[width=\columnwidth]{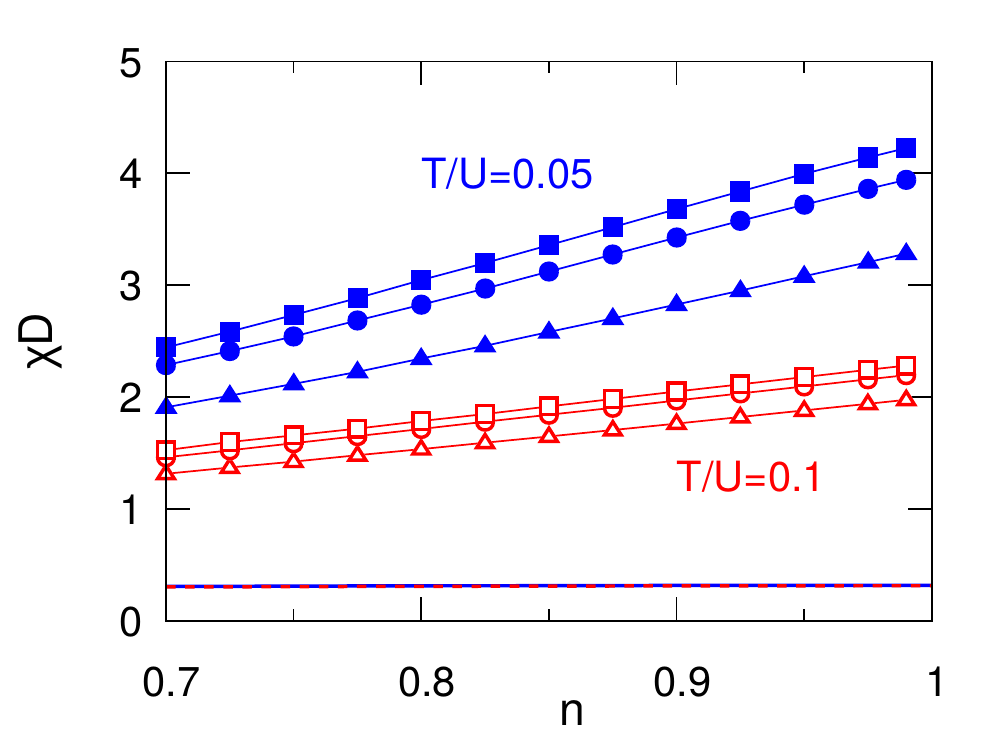}
    \caption{Magnetic Susceptibility as a function of band filling $n$
      in the system on the Bethe lattice
      when $U/D=10$ (triangles), $20$ (circles) and $30$ (squares)
      at the temperatures $T/D=0.1$ (open symbols) and $0.05$ (solid symbols).
      Solid and dashed lines around $\chi D\sim 0.3$ are the results
      for the noninteracting system at $T/D=0.05$ and $0.1$.
    }
    \label{fig:B-nchi}
  \end{center}
\end{figure}

\begin{figure}[htb]
  \begin{center}
    \includegraphics[width=\columnwidth]{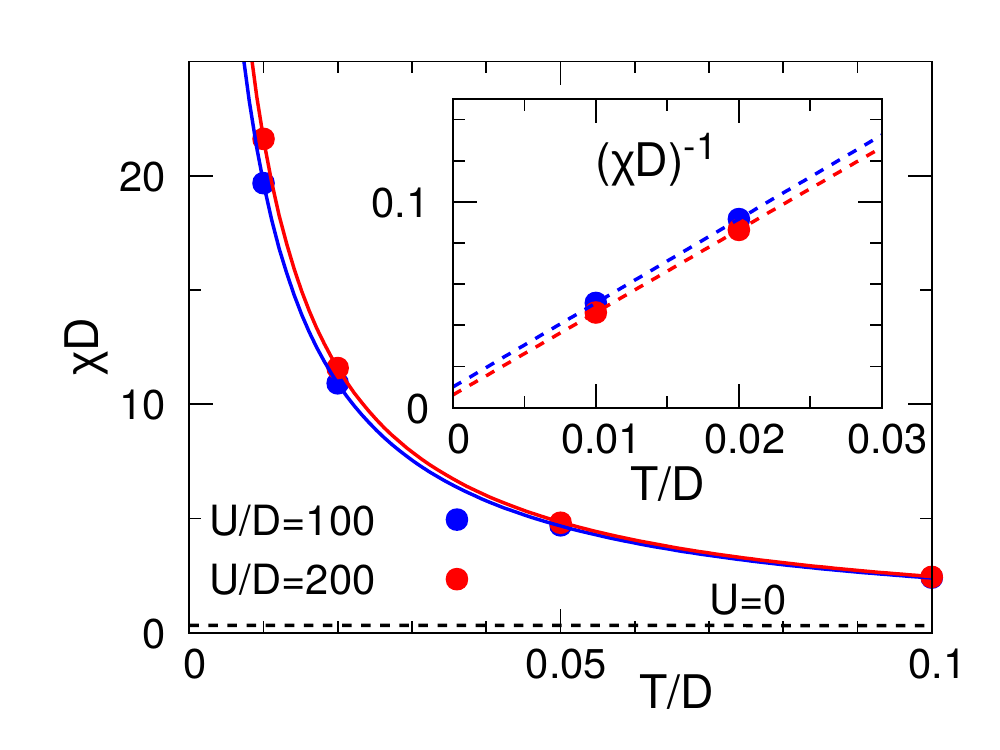}
    \caption{Magnetic susceptibility as a function of the temperature
      in the system on the Bethe lattice in the slightly doped systems $n=0.99$
      with $U/D=100$ and $200$.
      Inset shows the inverse of the susceptibility.
      Dashed lines are deduced from two data for $T/D=0.01$ and $0.02$.}
    \label{fig:B-Tchi}
  \end{center}
\end{figure}

%%%%%%%%%%%%%%%%%%%%%%%%%%%%%%%%%%%%%%
%\subsection{Bethe lattice}
%%%%%%%%%%%%%%%%%%%%%%%%%%%%%%%%%%%%%%
We also examine the ferromagnetism in the Hubbard model
on the Bethe lattice with the semielliptical DOS.
The results for the magnetic susceptibility at $T/D=0.05$ and $0.1$
are shown in Fig.~\ref{fig:B-nchi}.
%%%%%%%%%%%%%%%%%%%%%%%%%%%%%%%%%
%%%%%%%%%%%%%%%%%%%%%%%%%%%%%%%%%
We find that the susceptibility monotonically increases with increasing $n$
when the temperature and interaction strength are fixed.
This suggests that the magnetic instability should appear
in the vicinity of the half filling,
in contrast to the Hubbard model on the hypercubic lattice discussed above.
To examine whether or not the FM ordered state is realized at low temperatures,
we also calculate the temperature dependence of the susceptibility for the nearly half-filled system
($n=0.99$), as shown in Fig.~\ref{fig:B-Tchi}.
%%%%%%%%%%%%%%%%%%%%%%%%%%%%%%%%%
%%%%%%%%%%%%%%%%%%%%%%%%%%%%%%%%%
It is found that, in the system with the strong interactions
$U/D=100$ and $200$,
the magnetic susceptibility monotonically increases with decreasing temperatures.
However, we cannot find tendencies toward divergence
(see the inset of Fig.~\ref{fig:B-Tchi}).
This suggests the absence of the FM ordered state
in the single band Hubbard model on the Bethe lattice,
which is consistent with the previous works~\cite{Peters,BalzerPotthoff}.

Up to now, we have treated the hypercubic and Bethe lattices
to elucidate the origin of the magnetic instability to the FM ordered state
in the single-band Hubbard model; the detailed finite temperature calculations clarified that
the FM ordered phase appears in the hypercubic lattice,
while this does not in the Bethe lattice.
These facts might be understood by the Nagaoka mechanism~\cite{Nagaoka};
in the $U\rightarrow\infty$ limit,
the FM ordered state is realized in the one-hole doped half-filled system
with the closed-loop lattice structure.
However, in the framework of DMFT, the lattice structure is indirectly treated
only via the noninteracting DOS [Eq. (\ref{rho})].
Therefore, it may be difficult to conclude that the loop structure plays an essential role
in stabilizing the FM ordered state in the infinite dimensions.
Now, we focus on the DOS in the noninteracting system.
It is clear that the DOS around the Fermi level is similar to each other.
This suggests that the FM ordered state found in the present system with large interactions
is not attributed to the DOS at the Fermi energy,
which is crucial for the FM ordered state caused by the Slater mechanism justified
in the weak coupling limit.

\begin{figure}[htb]
  \begin{center}
    \includegraphics[width=\columnwidth]{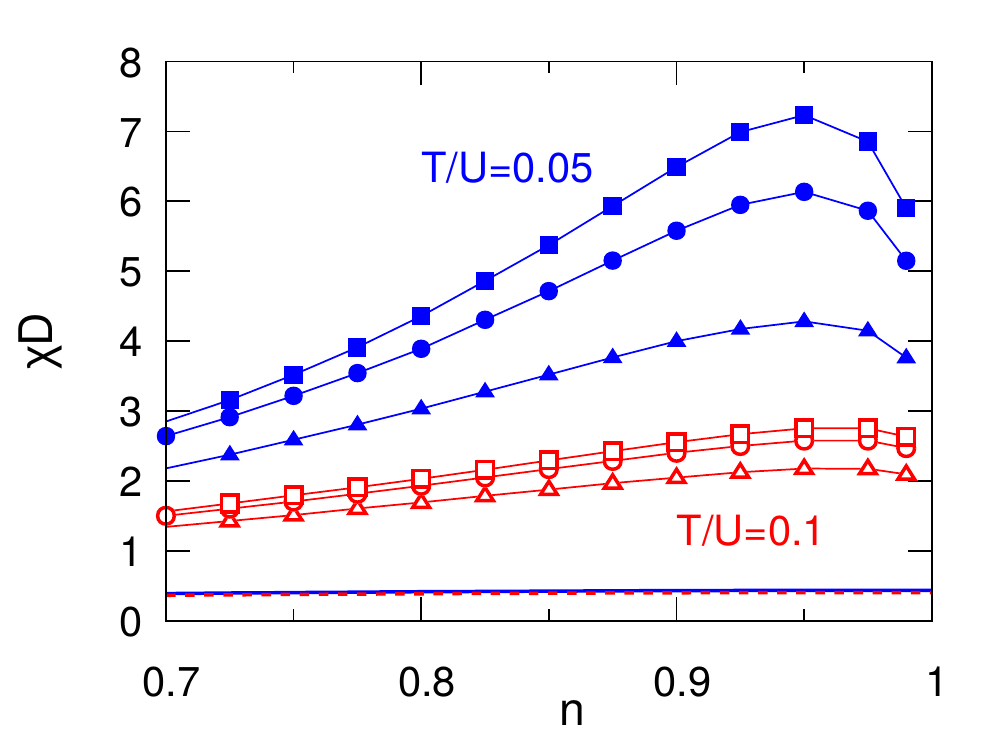}
    \caption{Magnetic Susceptibility as a function of band filling $n$
      in the system with $t$-distribution with $\nu=3$
      when $U/D=10$ (triangles), $20$ (circles) and $30$ (squares)
      at the temperatures $T/D=0.1$ (open symbols) and $0.05$ (solid symbols).
      Solid and dashed lines around $\chi D\sim 0.3$ are the results
      for the noninteracting system at $T/D=0.05$ and $0.1$.
    }
    \label{fig:student-nchi}
  \end{center}
\end{figure}

On the other hand, in the high energy region, there exists a clear difference in DOS;
$\rho_b=0$ for the Bethe lattice, while $\rho_{hc}\neq 0$ for the hypercubic lattice.
This expects that the asymptotic form of DOS away from the Fermi level
plays an important role in stabilizing the FM ordered state in the strong-coupling limit.
Here, 
we introduce another function form, so-called $t$-distribution~\cite{student},
\begin{eqnarray}
  \rho_t (x , \nu) &=& \frac{\Gamma \left( \displaystyle \frac{\nu + 1}{2} \right)}{\sqrt{\pi\nu} \Gamma \left(\displaystyle\frac{\nu}{2}\right)} \left[1 + \frac{1}{\nu}x^2\right]^{-\frac{\nu + 1}{2}},
\end{eqnarray}
%%%%%%%%%%%%%%%%%%%%%%%%%%%%%%%%%
where $\Gamma(x)$ is the Gamma function.
%Since the DOS decays with $|x|^{-(\nu+1)}$ $(|x|\rightarrow\infty)$,
%it is always finite.
This is reduced to the Cauchy-Lorentz distribution in the case $\nu=1$ and
the Gaussian distribution (hypercubic) in the case $\nu \to \infty$.
As an example, we consider the $t$-distribution with $\nu=3$,
%%%%%%%%%%%%%%%%%%%%%%%%%%%%%%%%%%
\begin{align}
  \rho_t (x , 3) = \frac{1}{\sqrt{2}\pi D} \left[\left(\frac{x}{D}\right)^2 + \frac{1}{2}\right]^{-2},
  \label{eq:t}
\end{align}
where $D$ is the unit of energy, which is determined such that
its variance coincides with that of the DOS of the hypercubic lattice given in Eq.~(\ref{eq:1}).
The filling dependence of the magnetic susceptibility
in the system with $\rho_t (x , 3)$ is shown in Fig.~\ref{fig:student-nchi}.
%%%%%%%%%%%%%%%%%%%%%%%%%%%%%%%%%
%%%%%%%%%%%%%%%%%%%%%%%%%%%%%%%%%
We find that nonmonotonic behavior appears in the magnetic susceptibility and
the maximum of the curves is located around $n\sim 0.95$ at $T/D=0.1$.
The results are similar to those for the hypercubic lattice,
which expects that the FM ordered state is realized in a finite parameter space
unlike the case on the Bethe lattice.
%%%%%%%%%%%%%%%%%%%%%%%%%%%%%%%%%
\begin{figure}[htb]
  \begin{center}
    \includegraphics[width=\columnwidth]{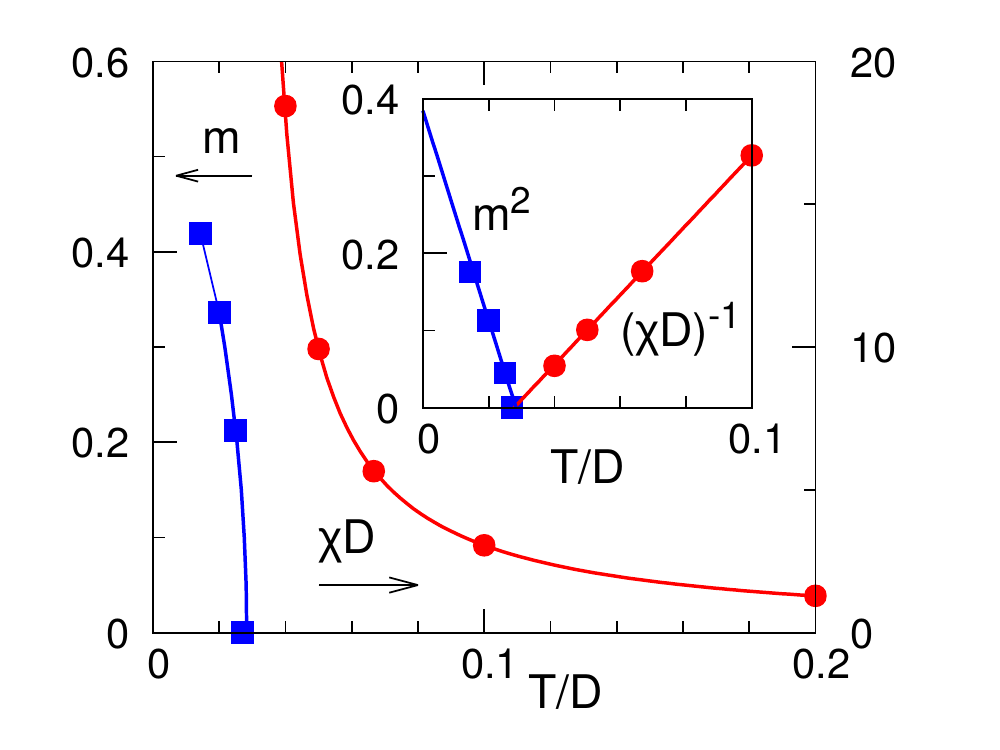}
    \caption{Magnetization and magnetic susceptibility as a function of the temperature
      in the system with $t$-distribution when $U/D=100$ and $n=0.95$.
      The inset shows critical behavior of the susceptibility and magnetization.
      Solid lines are guides to the eyes.
}
    \label{fig:student-Tchiinv}
  \end{center}
\end{figure}
%%%%%%%%%%%%%%%%%%%%%%%%%%%%%%%%%
Figure~\ref{fig:student-Tchiinv} shows the magnetization and susceptibility
in the system with $U/D=100$ and $n\sim 0.95$.
Decreasing temperatures, the susceptibility diverges
at the critical temperature $T_c/D=0.028$ and
the magnetization appears below the temperature.
This temperature is larger than that in the case with the Gaussian DOS,
indicating that the FM ordered state in the system with the $t$ distribution DOS is
more stable than that in the hypercubic system in the case with the strong Coulomb interaction.

%%%%%%%%%%%%%%%%%%%%%%%%%%%%%%%%%%%%%%
\section{Summary}\label{sec5}
%%%%%%%%%%%%%%%%%%%%%%%%%%%%%%%%%%%%%%
We have studied magnetic properties in the single-band Hubbard model in the infinite dimensions.
Combining DMFT with the continuous-time quantum Monte Carlo simulations,
we have calculated uniform magnetic susceptibility and magnetization systematically and
have found that the FM ordered state is realized in the system on the hypercubic lattice,
while no ordered state appears on the Bethe lattice.
We have also examined the system with $t$-distribution DOS
which has a power-law tail unlike the Gaussian distribution.
We have found that the FM ordered state in the system with $t$-distribution is more
stable than with that in the system on the hypercubic lattice.

The present results suggest that
the noninteracting DOS in the high energy region
contributes to the stability of the FM ordered state in the strong-coupling regime
while the DOS
around the Fermi level is not relevant to the emergence of the FM ordered phase.
This is in contrast to the Stoner ferromagnetism in the weak coupling limit,
where the DOS at the Fermi energy is crucial for the emergence of the ferromagnetism.
It is then expected that, in the finite dimensional systems on a simple lattice
such as square and cubic lattices,
no FM instability appears due to the absence of DOS in the high energy region.
It is an interesting problem to discuss the the FM instability,
by taking into account non-local electron correlations,
which is now under consideration.

\begin{acknowledgments}
We would like to thank J. Otsuki, H. Shinaoka, and P. Werner for valuable comments.
Parts of the numerical calculations were performed
in the supercomputing systems in ISSP, the University of Tokyo.
This work was supported by Grant-in-Aid for Scientific Research from
JSPS, KAKENHI Grant Nos. JP18K04678, JP17K05536 (A.K.),
JP16K17747, JP16H02206, JP18H04223 (J.N.).
The simulations have been performed using some of the ALPS libraries~\cite{alps2}.
\end{acknowledgments}

\bibliography{./refs}

\end{document}